\documentclass[epj]{svjour}

\usepackage{bbm}
\usepackage{amsfonts}
\usepackage{latexsym}
\usepackage{amsmath}
\usepackage{amssymb}
\usepackage{bbm}
\usepackage{color}

\usepackage{graphicx}
\usepackage{fancyhdr}
\def\makeheadbox{{
 }}
\def\mytitle{My title} 
\def\myauthors{My name}  
\def\mytype{My type of session}
\def\mysession{My session}

\def\mytitle{Blowups of Heterotic Orbifolds using Toric Geometry} 
\def\myauthors{Stefan Groot Nibbelink}    
\def\mytype{Contributed Talk}    
\def\mysession{Theoretical Models}

\renewcommand{\d}{\mathrm{d}}

\newcommand{\captn}[1]{\vspace{-3ex}\caption{\small #1}}

\DeclareMathSymbol{\mg}{\mathrel}{symbols}{"1D}

\renewcommand{\ge}{\epsilon}

\newcommand{\gf}{\phi}

\newcommand{\gth}{\theta}

\newcommand{\gp}{\pi}

\newcommand{\cF}{{\cal F}}

\newcommand{\cK}{{\cal K}}

\newcommand{\cR}{{\cal R}}

\newcommand{\tZ}{{\tilde Z}}
\newcommand{\tr}{\text{tr}}

\newcommand{\ra}{\rightarrow}

\newcommand{\dsp}{\displaystyle}

\newcommand{\labl}[1]{\label{#1}}
\newcommand{\Kh}{K\"{a}hler}
\newcommand{\beq}{\begin{equation}}
\newcommand{\eeq}{\end{equation}}
\newcommand{\barr}{\begin{array}}
\newcommand{\earr}{\end{array}}
\newcommand{\equ}[1]{\begin{gather} #1 \end{gather}}

\newcommand{\pmtrx}[1]{\begin{pmatrix} #1 \end{pmatrix}}

\newcommand{\non}{\nonumber}
\newcounter{oldcounter}

\newcommand{\bee}{{\bar e}}

\newcommand{\bz}{{\bar z}}
\newcommand{\bge}{{\bar\epsilon}}

\newcommand{\Intr}{\mathbb{Z}}
\newcommand{\Cplx}{\mathbb{C}}

\newcommand{\CP}{\mathbb{CP}}

\newcommand{\ba}[2]{\[\begin{array}{#2}\label{#1}}
\newcommand{\ea}{\end{array}\]}
\newcommand{\be}{\begin{equation}}
\newcommand{\ee}{\end{equation}}
\newcommand{\bea}{\begin{eqnarray}}
\newcommand{\eea}{\end{eqnarray}}

\newcommand{\U}[1]{\mathrm{U(#1)}}
\newcommand{\SU}[1]{\mathrm{SU(#1)}}
\newcommand{\SO}[1]{\mathrm{SO(#1)}}

\newcommand{\rep}[1]{\mathbf{#1}}
\newcommand{\crep}[1]{\overline{\rep{#1}}}

\newcommand{\sm}{{\,\mbox{-}}}

\def\fgregdelta{
\begin{figure}
\begin{center} 
{\includegraphics[width=.44\textwidth,clip=true,viewport=0 30 370 226]{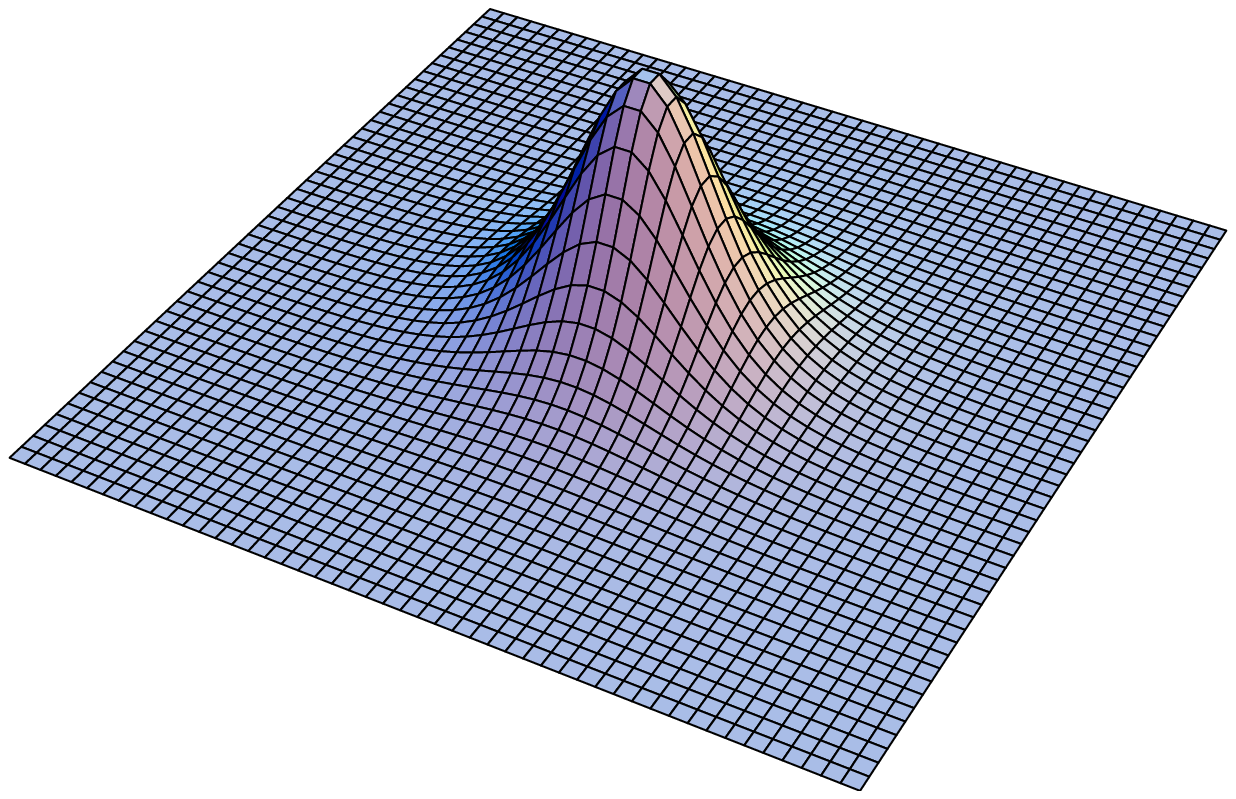}}
\end{center}
\caption{\label{fg:regdelta}
The curvature~\eqref{Curv2expl} mimics a
regularized delta--function.  
}
\end{figure}
}

\def\modelsCthreeZfour{
\begin{table*}
\caption{\label{tb:modelsC3Z4}
This table compares the $\Cplx^3/\Intr_4$ orbifold gauge shift vector $v\,$, 
with the blowup vectors $V_1$ and $V_2\,$,
that characterize the line bundle gauge background on the resolution. 
}
\begin{center}
\begin{tabular}{c c}
\begin{tabular}[t]{| c | c  c | l |}
\hline &&& \\ [-2ex]
orbifold & blowup & blowup &
\\ [-1ex] &&& \\[-2ex]
shift $4\,v$ & vector $V_2$ & vector $V_1$  & Nr. 
\\[1ex]\hline\hline &&& \\ [-2ex]
$(0^{13},1^2,2)$ 
& $(0^{13},1^2,2)$ 
& $(0^{13},1^2,$-$2)$ 
& 1a
\\[.7ex]
& $(0^{13},1^2,2)$
& $(0^{12},2,$-$1^2,0)$
& 1b
\\[.7ex]
& $(0^{13},1^2,2)$
& $(0^{11},2,1,0^2,$-$1)$
& 1c  
\\[1ex]\hline &&& \\ [-2ex]
$(0^{11},1^2,2^3)$ 
& $(0^{13},1^2,2)$ 
& $(0^{10},1^4,$-$1^2)$
& 2a 
\\[.7ex]
&  $(0^{13},1^2,2)$
&  $(0^{11},1^2,$-$2,0^2)$
& 2b 
\\[1ex]\hline &&& \\ [-2ex]
$(0^{9},1^2,2^5)$ 
& $(0^{13},1^2,2)$
& $(0^{8},1^5,0^2,$-$1)$
& 3a 
\\[.7ex]
&  $(0^{13},1^2,2)$
&  $(0^{9},1^4,$-$1^2,0)$
& 3b  
\\[1ex]\hline &&& \\ [-2ex]
$(0^{7},1^2,2^7)$ & $-$ & $-$ & 4 
\\[1ex]\hline &&& \\ [-2ex]
$(0^{10},1^6)$ 
& $(0^{10},1^6)$ 
& $(0^{10},1^2,$-$1^4)$ 
& 5a
\\[.7ex]
&  $(0^{10},1^6)$
&  $(0^{13},1,$-$1,$-$2)$
& 5b  
\\[1ex]\hline &&& \\ [-2ex]
$(0^{10},1^5,3)$ 
& $(0^{10},1^6)$
& $(0^{9},2,$-$1^2,0^4)$
& 6 
\\[1ex]\hline &&& \\ [-2ex]
$(0^{8},1^6,2^2)$ 
& $(0^{10},1^6)$
& $(0^{8},1^3,$-$1^3,0^2)$
& 7a 
\\[.7ex]
&  $(0^{10},1^6)$
&  $(0^{8},1^2,$-$2,0^5)$
& 7b  
\\[1ex]\hline &&& \\ [-2ex]
$(0^{6},1^6,2^4)$ 
& $(0^{10},1^6)$
& $(0^{6},1^4,$-$1^2,0^4)$
& 8 
\\[1ex]\hline
\end{tabular} \hspace{-14pt} &
\begin{tabular}[t]{| c | c c | c |}
\hline &&& \\ [-2ex]
orbifold & blowup & blowup &
\\ [-1ex] &&& \\[-2ex]
shift $4\,v$ & vector $V_2$ & vector $V_1$  & Nr. 
\\[1ex]\hline\hline &&& \\ [-2ex]
$(0^{5},1^{10},2)$ 
&  $(0^{10},1^6)$
& $\frac{1}{2}($-$3,1^{10},$-$1^5)$
& 9 
\\[1ex]\hline &&& \\ [-2ex]
$(0^{3},1^{10},2^3)$ 
& $(0^{10},1^6)$
& $\frac{1}{2}(1^{12},$-$1^3,$-$3)$
& 10 
\\[1ex]\hline &&& \\ [-2ex]
$(1^{14},2^2)$ 
& $(0^{13},$-$2,1^2)$
& $\frac{1}{2}(1^{15},$-$3)$
& 11 
\\[1ex]\hline &&& \\ [-2ex]
$(1^{13},$-$1,2^2)$ 
& $(0^{13},1^2,2)$
& $\frac{1}{2}(1^{15},$-$3)$
& 12a 
\\[.7ex]
& $(0^{13},1^2,2)$
& -$\frac{1}{2}($-$3,1^{15})$
& 12b 
\\[1ex]\hline &&& \\ [-2ex]
$\frac{1}{2}(1^{3},3^{12},$-$3)$ 
& $\frac{1}{2}($-$3,1^{15})$
& $$-$(0^{13},1^2,2)$
& 13a 
\\[.7ex]
& $\frac{1}{2}(1^{15},$-$3)$
& $(0^{13},1^2,2)$
& 13b 
\\[.7ex]
& $\frac{1}{2}(1^{15},$-$3)$
& $\frac{1}{2}(1^3,$-$1^{11},3,1)$
& 13c 
\\[1ex]\hline &&& \\ [-2ex]
$\frac{1}{2}(1^{7},3^{8},$-$3)$ 
& $\frac{1}{2}(1^{15},$-$3)$
& $($-$1^5,1,0^{10})$
& 14a 
\\[.7ex]
& $\frac{1}{2}(1^{15},$-$3)$
& $\frac{1}{2}(1^{6},$-$1^8,$-$3,1)$
& 14b 
\\[1ex]\hline &&& \\ [-2ex]
$\frac{1}{2}(1^{11},3^{4},$-$3)$ 
& $\frac{1}{2}(1^{15},$-$3)$
& $(0^{10},1^3,$-$1^3)$
& 15 
\\[1ex]\hline &&& \\ [-2ex]
$\frac{1}{2}(1^{15},$-$3)$ 
& $\frac{1}{2}(1^{15},$-$3)$
& $(0^{13},$-$2,1^2)$
& 16a 
\\[.7ex]
& $\frac{1}{2}(1^{15},$-$3)$
& $\frac{1}{2}($-$1^{14},3,$-$1)$
& 16b 
\\[1ex]\hline 
\end{tabular}
\end{tabular}
\end{center}
\end{table*}
}

\def\tbZthreeorMod{
\begin{table*}
\caption{\label{tb:Z3orMod}
The first column displays the heterotic $\Intr_3$ $\SO{32}$ orbifold
shifts. The $\U{1}$ bundles on the
blowup are defined by the second column. The gauge groups of the
heterotic orbifold models are listed in the next column. The one but
last column contains the matter representations on the resolution. The
last column gives the  {\em additional} twisted matter. 
}\begin{center} 
\begin{tabular}{| c | c | c | c | c |}
\hline &&&& \\ [-2ex]
Orbifold  & Blowup & $G_{\rm orbifold}=$ & Matter spectrum on the  & {Additional}  \\ [0ex]
shift & shift & $G_{\rm blow\ down}$ 
& orbifold resolution & twisted matter 
\\[1ex]\hline &&&&\\ [-2ex]
$(0^{13},1^2,2)$  & $(0^{12},1^3,3)$ & 
$\SO{26} \times \U{3}$ & 
$\frac 19 (\rep{26},\rep{3}) + \frac{26}{9} (\rep{1},\crep{3})  
+ (\rep{26},\rep{1})$& $(\rep{1},\rep{1})$
\\[1ex] &&&&\\ [-2ex]
& $(0^{13},2^3)$  & 
&
$\frac 19 (\rep{26},\crep{3}) + \frac{26}{9} (\rep{1},\rep{3})$
& $(\rep{1},\rep{1})+(\rep{26},\rep{1})$
\\[1ex]\hline &&&&\\ [-2ex]
 $(0^{10},1^4,2^2)$ & $(0^{10},1^4,2^2)$ & 
$\SO{20} \times \U{6}$ &
$ \frac {10}9 (\rep{1},\crep{15}) 
+ \frac 19 (\rep{20},\rep{6}) 
+ 3 (\rep{1}, \rep{1})$& 
\\[1ex]\hline &&&& \\ [-2ex]
$(0^{7},1^6,2^3)$ & $(0^{7},1^8,2)$ & 
$\SO{14} \times \U{9}$ &
$  \frac 19 (\rep{14}, \rep{9})
+ \frac 19 (\rep{1},\crep{36})
+ (\rep{1}, \crep{9}) $ & 
\\[1ex]\hline &&&& \\ [-2ex]
$(0^{4},1^8,2^4)$ & $(0^{4},1^{12})$ & 
$\SO{8} \times \U{12}$ &
$ \frac 19 (\rep{8}, \rep{12})
+ \frac 19 (\rep{1}, \crep{66})$&
$(\rep{1},\rep{1})+(\rep{8_+},\rep{1})$
\\[1ex] &&&& \\ [-2ex]
& $(\frac 12^{12},\frac 32^{4})$ & 
&
$ \frac 19 (\rep{8}, \crep{12})
+ \frac 19 (\rep{1}, \rep{66})+(\rep{8_+},\rep{1})$
& $(\rep{1},\rep{1})$
\\[1ex]\hline &&&& \\ [-2ex]
$(0^{1},1^{10},2^5)$ & $\!(\frac 12^{14},\frac 32, \sm\frac 52)$ & 
$\SO{2}\times \U{15}$ &
$ \frac {11}9 (\rep{15})  
+ \frac 19 (\crep{105})+ 3(\rep{1}) $ 
&
\\[1ex]\hline 
\end{tabular} 
\end{center}
\end{table*}
}

\begin{document}
\title{Blowups of Heterotic Orbifolds using Toric Geometry}
\author{Stefan Groot Nibbelink\inst{1,2}
}                     
%
%
\institute{
Institute for Theoretical Physics, 
University Heidelberg, 
Philosophenweg 19, 
D-69120 Heidelberg,
Germany
\and 
Shanghai Institute for Advanced Study, 
USTC,
99 Xiupu Rd, Pudong, Shanghai 201315, P.R.\ China
}
%
\date{August 15, 2007}
\abstract{
Heterotic orbifold models are promising candidates for models with
MSSM like spectra. But orbifolds only correspond to a special place in
moduli space, the bigger picture is described by the moduli space of
Calabi-Yau spaces. In this talk we will make explicit connections
between both points of view. To this end we study blowups of orbifold
singularities using both explicit constructions and
toric geometry techniques. We show that matching of all orbifold
models in blowups are possible.
} 
\maketitle
\section{Introduction and summary}
\label{sc:intro}

One of the central aims of string phenomenology is to build string models
reproducing  the supersymmetric standard model of particle physics. 
There have been various approaches in this direction: 
Free--fermion models~\cite{Faraggi:1989ka,Faraggi:1991jr}, 
intersecting D--branes in type II string theory~\cite{Berkooz:1996km,Blumenhagen:2000wh,Aldazabal:2000dg,Honecker:2004kb},
Gepner models~\cite{Dijkstra:2004cc,Dijkstra:2004ym}, 
and compactifications of the heterotic string. In the latter case in
order to obtain at most four dimensional $N=1$ supersymmetry one needs
to compactify on a Calabi--Yau  space~\cite{Candelas:1985en} (for
recent progresses 
see~\cite{Andreas:1999ty,Braun:2005bw,Blumenhagen:2005zg,Blumenhagen:2006ux}).
Orbifolds (singular limits of Calabi--Yaus) are 
convenient, because they allow for calculable string
compactifications~\cite{dixon_85,Dixon:1986jc}. It is
possible to produce a vast but controllable landscape of models, and
scan among them for realistic ones. Indeed, this approach has been
proven to be successful, and models close to the MSSM have
been
built~\cite{Forste:2004ie,Kobayashi:2004ya,Kobayashi:2004ud,Buchmuller:2004hv,Lebedev:2006kn,Kim:2006hw}.

Orbifolds are special points in the full moduli space of the heterotic
string on Calabi--Yau manifolds. In order to have control on
the theory away from these special points, it is crucial to
have a better understanding of model building on the corresponding
smooth compactification spaces. 
A concrete way to probe the moduli space surrounding 
orbifold points is to consider blowups of
orbifold singularities. 
The construction of explicit blowups is unfortunately
not easy. The best known example is the Eguchi--Hanson
resolution~\cite{Eguchi:1978xp} of the $\Cplx^2/\Intr_2$ orbifold
singularity. Generalization to $\Cplx^n/\Intr_n$ was discussed
in~\cite{Calabi:1979}.  
The singularities of more complicated orbifolds might not
allow for a simple explicit blowup construction. On the other
hand, the topological properties of such resolutions can be
conveniently described by toric geometry, see
e.g.~\cite{Erler:1992ki}.

In this talk we explain how using both explicit blowups and toric
geometry one can construct heterotic models on orbifold resolutions:   
We construct explicit {blowups} of {$\Cplx^n/\Intr_n$ orbifolds} with
{U(1) gauge bundles}~\cite{Ganor:2002ae,Nibbelink:2007rd}. 
We compare the resulting {spectra} with that of {heterotic
$\Cplx^3/\Intr_3$ orbifolds}. We  {reproduce most} 
of the {twisted states}; the ``missing'' states either got mass or are
reinterpreted as non--universal axions. (Multiple anomalous U(1) gauge
fields in
{blowup} are possible~\cite{Blumenhagen:2005pm}: 
Anomalous {field redefinitions} avoid contradictions 
with the orbifold picture with at most a single anomalous
U(1)~\cite{GrootNibbelink:2007ew}.) 
Finally, in this talk we show that {similar analysis} on {more
complicated orbifolds}, like $\Cplx^3/\Intr_4$, is {doable}. 
Applications to resolutions of other orbifolds, such as 
{$\Cplx^2/\Intr_3$} and {$\Cplx^3/\Intr_2\times \Intr_2'$} can be
found in~\cite{Nibbelink:2007pn}. We obtain exact agreement between
{blowup} and {heterotic orbifold spectra on $\Cplx^2/\Intr_2$},
consistent with~\cite{Honecker:2006qz}. In future work we 
investigate resolutions of the phenomenological interesting
$\Intr_{6-II}$ orbifolds.

\section{Explicit blowup of $\boldsymbol{\Cplx^n/\Intr_n}$ singularity}
\label{sc:explicit}

We review the explicitly construction of a blowup of the
$\Cplx^n/\Intr_n$ orbifold with possible $\U{1}$ bundles
following~\cite{Ganor:2002ae,Nibbelink:2007rd}. The $\Cplx^n/\Intr_n$ 
orbifold is defined by the $\Intr_n$ action 
$\tZ \ra \gth \, \tZ\,$, where $\gth = e^{2\pi i\, \gf}\,,$ with 
$\gf = ( 1, \ldots, 1 )/n\,$.
The geometry of the non--singular blowup is described
by the \Kh\ potential
\equ{
\cK(X) ~=~ 
\int\limits_1^X \frac {\d X'}{X'} \, 
 \frac 1{n} \big( r + X \big)^{\frac 1n}~, 
\labl{FunK}
}
where $X = (1 + \bz z)^n |x|^2$ is an $\SU{\mbox{$n$}}$ invariant, and
the $z$ and $x$ are the coordinates of the space. The resolution
parameter $r$ is defined such that in the limit $r \ra 0$ one
retrieves the orbifold geometry.

\fgregdelta

From the \Kh\ potential all geometrical quantities can be derived in
the standard way, in particular, the curvature 2--form reads 
\equ{
\cR = \frac{r}{r + X} \! 
\pmtrx{ \dsp 
 e \, \bee - \bee \, e  
+ \frac 1{n}\, \frac{ \bge \, \ge}{r + X}
~\quad 
\frac{\bge \, e} {\sqrt{r + X}} 
\\[3ex] \dsp 
\frac{\bee\, \ge}{\sqrt{r + X}} 
~\quad 
n\, \bee \, e - \frac {n\!-\!1}{n} \,  \frac{ \bge \, \ge}{r + X}
}.
\labl{Curv2expl}
}
Here $e$ and $\ge$ are the holomorphic vielbein 1--forms of
$\CP^{n-1}$ and its complex line bundle. An impression of the
curvature is given in figure~\ref{fg:regdelta}. This
geometry admits a $\U{1}$ gauge background satisfying the
Hermitian Yang--Mills equations  
\equ{
i\cF_V = \Big(\frac r{r + X} \Big)^{1-\frac 1n}
\Big( 
\bee e - \frac {n-1}{n^2} \, \frac 1{r+X}\, \bge \ge
\Big) H_V~, 
\labl{FU1basis}
}
where $H_V = V^I H_I$ with $H_I$ Cartan generator and $V^I$ either all
integers or half integers. 
Because both the geometry and its $\U{1}$ gauge background are given
explicitly, integrals of them can be computed:
\equ{ 
\int_{\CP^2} \frac{\tr\, \cR^2}{(2\gp i)^2} 
= -n \, 
 \int_{\CP^1\ltimes \Cplx}  \frac {\tr\, \cR^2}{(2\gp i)^2}
= 
n(n+1)~, 
\labl{trR2int}
\\[1ex] 
\int_{\CP^p} 
\Big( \frac { i \cF}{2\gp i} \Big)^p
~=~ 
- n
\int_{\CP^{p-1} \ltimes \Cplx} 
\Big( \frac { i \cF}{2\gp i} \Big)^p
~=~ 
1~. 
\labl{trFpint}
}
The integrals over $\CP^{p}$ are taken at $X=0$ integrating over $p$
of the $n-1$ inhomogeneous coordinates of $\CP^{n-1}$. The integral over 
$\CP^{p-1} \ltimes \Cplx$ corresponds to the integral over all values
of  $x \in \Cplx$ and over $p-1$ inhomogeneous coordinates.

Using the explicit geometry of the blowup of $\Cplx^3/\Intr_3$ with
$\U{1}$ gauge bundle, we can construct string compactifications. 
The integrated Bianchi identity integrated over $\CP^2$ has
to vanish, giving:  $V^2 = 12.$
The same condition is found when integrating over $\CP^1\ltimes
\Cplx$ and selects 7 allowed models listed in table~\ref{tb:Z3orMod}.   
The spectra of these models can be compute using an index
  theorem. The multiplicities of the representations obtained
from the branching of the adjoint of $\SO{32}$ via the multiplicity 
operator $N_V$ which can take the values:  
$N_V = \frac 19,~ 1,~ \frac {26}9 = 3 - \frac 19$. 
The multiplicity factor $\frac 19 = \frac{3}{27}$ refers to
untwisted (delocalized) states, while integral multiplicity
  factors correspond to states localized at the orbifold fixed
point~\cite{Gmeiner:2002es}. The table~\ref{tb:Z3orMod} compares the
matter on the blowup with the heterotic orbifold spectrum in the blow
down limit, and shows that only sometimes some vector--like matter is
not recovered on the blowup.

\tbZthreeorMod

\section{Toric resolutions of orbifold singularities}

\begin{figure}
\begin{center} 
\begin{tabular}{ c c c  }
\raisebox{0ex}{\scalebox{0.4}{\mbox{\input{C3Z3.pstex_t}}}}
&\qquad & 
\raisebox{0ex}{\scalebox{0.4}{\mbox{\input{C3Z4.pstex_t}}}}

\end{tabular} 
\end{center}
\captn{\label{fg:toricC3Z3}
Projected views
of the toric diagrams of the resolutions of $\Cplx^3/\Intr_3$ (left)
and $\Cplx^3/\Intr_4$ (right). 
}
\end{figure}

We do not have the time to explain the
properties of toric geometry~\cite{Fulton,Hori:2003ic,Lust:2006zh} in
detail. The rough idea of toric resolutions of 
orbifold singularities is to replace the orbifold action by
invariance $\Cplx^*$ scalings of the coordinates $z_i$. To
keep the dimensionality of the resolution equal to that of the
orbifold one needs to introduce as many extra coordinates $x_p$ as
complex scalings. Setting one of the homogeneous coordinates of the
resolution defines a codimension one hypersurface called 
a divisor. Ordinary divisors are
defined by $D_i = \{ z_i = 0 \}$, and exceptional divisors by
$E_p = \{ x_p = 0 \}$. To each divisor we can associate a
line bundle characterized by the transition functions between the
various coordinate patches of the defining equation of the divisor.
 The first Chern class of a line bundle is a $(1,1)$--form,
and hence we can reinterpret the divisors as $(1,1)$--forms
themselves. Not all divisors are independent because of so--called
linear equivalence relations among them 
\equ{
\sum_i (v_i)_j \, D_i ~+~ \sum_p (w_p)_j \, E_p ~\sim~ 0~. 
}
As there are as many such linear equivalence relations as ordinary
divisors, we may take the exceptional divisors as a
basis for the gauge background $\cF_V$.

As hypersurfaces the divisors can intersect multiple times. These
intersection numbers can be reinterpreted as integrals of the
corresponding $(1,1)$--forms over the whole resolution. The
intersections define the complete topology of the resolution. This
topological information is conveniently summarized in the 
toric diagram: In a toric diagram the divisors are denoted as nodes,
curves i.e.\ intersection of two divisors as lines between two nodes,
and intersections of three different divisors as cones spanned by
three nodes. Basic cones, the smallest possible cones, define
intersections of three divisors with unit intersection number, while
lines of three nodes correspond to intersection number zero.  
Together with the linear equivalence relations the toric diagram
determines all (self--)intersections.

\section{Toric resolution of $\boldsymbol{\Cplx^3/\Intr_3}$}

We illustrate the power of toric geometry by reproducing the results
obtained using the explicit blowup of $\Cplx^3/\Intr_3$. The toric
resolution of this orbifold has three ordinary divisors
$D_i$, and a {single exception one $E$}. 
They satisfy the {linear equivalence relations}: 
\equ{
{D_i ~\sim D_j}~, 
\qquad 
{3\, D_i} \,+\, {E} ~\sim~ 0~, 
\labl{lineqvCnZn}
}
From the {toric diagram}, left picture in figure~\ref{fg:toricC3Z3},
we infer the basic integrals and intersections:  
${D_1 D_2 E } = {D_2 D_3 E } = { D_3 D_1 E } = 1.$
The gauge field strength can be expanded as 
$\cF_V = -\frac 13 \, E\, H_V$. We obtained {all the results} of 
the {explicit blowup}. In particular, the
{Bianchi identity} on the {compact} cycle $E$ gives:
\equ{
V^2 = \int_{E} \tr (i\cF_V)^2 = \int_{E} \tr\, \cR^2 = 12~. 
}
The non--compact Bianchi identity follows immediately upon using the
linear equivalence relation~\eqref{lineqvCnZn} and leads to the same
condition.

\section{Heterotic models on resolution of  $\boldsymbol{\Cplx^3/\Intr_4}$}

The main advantage of using toric geometry over explicit blowups lies
in the fact that one can still use toric techniques in cases where no
explicit blowup is known. To exemplify this we consider the resolution
of $\Cplx^3/\Intr_4$. In this case there are two
exceptional divisors $E_1$ and $E_2$, which satisfy the linear
equivalence relations
\equ{
4\, D_1 \,+\, E_1 \,+\, 2\, E_2 ~\sim~ 0~, 
\quad 
4\, D_2 \,+\, E_1 \,+\, 2\, E_2 ~\sim~ 0~, 
\non \\[1ex]
2\, D_3 \,+\, E_1 ~\sim~ 0~. 
\labl{lineqvC3Z4}
}
To define the integrals on the resolution of $\Cplx^3/\Intr_4$ we use
the toric diagram, on the right hand side of
figure~\ref{fg:toricC3Z3}, and obtain  
\equ{ 
{D_1\, E_1\, E_2} = {D_2\, E_1\, E_2} = 
{D_1\, D_3\, E_1} = {D_2\, D_3\, E_1} = 1~,  
\non \\[1ex] 
D_1\, D_2\, E_2 = D_3\, E_1\, E_2 = 0~.
}
Via the linear equivalences this implies:
\equ{
E_1^2\, E_2  = 0~,~
E_2^2\, E_1 = -2~,~ 
E_1^3 = 8~,~
E_2^3 = 2~. 
}
The bottom {edge} of the toric diagram defines the {toric diagram}
of  the resolution of $\Cplx^2/\Intr_2$. The gauge background is
expanded in terms of the exceptional divisors
\equ{
{{\cF_V} = -\frac 12\, E_1\, H_1 
\,-\, \frac 14\, (E_1 + 2 \, E_2) H_2~,}
}
where $H_1 = V_1^I H_I$, etc. 
In order to ensure that we can directly compute the spectrum on the
non--compact resolution, we require that all the Bianchi identities
vanish on $E_1$, $E_2$ and the resolution of $\Cplx^2/\Intr_2$: 
\equ{
E_1:~~
{V_1^2 \,+\, V_1\cdot V_2 ~=~ 4}~, 
\quad 
E_2:~~ {V_1\cdot V_2 ~=~ -2~,} 
\non \\[1ex] 
\text{Res}(\Cplx^2/\Intr_2):~~
{V_2^2 ~=~ 6~.}  
}
The matching between the heterotic orbifold models and the resolution
models characterized by the shifts $V_1$ and $V_2$ is performed in
table~\ref{tb:modelsC3Z4}. All models except number 4 is reproduced in
blowup. This model is not obtained because it does not have any first
twisted sector, 
hence simply cannot be blown up. We have computed the complete
spectrum and confirmed that all blowup models have anomaly free
spectra~\cite{Nibbelink:2007pn}.

\modelsCthreeZfour

%
 \bibliographystyle{letter}
 \bibliography{paper}

\end{document}